\documentclass[prl,twocolumn,showpacs,showkeys]{revtex4-1}

\usepackage{latexsym,graphicx,graphics}
\usepackage{appendix}
\usepackage{amsmath,amssymb,epsfig}
\usepackage{float}

\newcommand{\abs}[1]{\left| #1 \right|}

\newcommand{\bb}[1]{ \mbox{\boldmath$ #1$}}

\newcommand{\rv}{\bb r}

\newcommand{\Eq}[1]{Eq.~(\ref{#1})}

\newcommand{\figstyle}[1]{\small{#1}}

\newcommand{\unit}[1]{\bb{\hat{#1}}}

\begin{document}

\title{Meta-Weaves: Sector-way non-reciprocal meta surfaces}
\author{Y. Mazor}
\surname{Mazor}
\author{Ben Z. Steinberg}
\surname{Steinberg}
\email{steinber@eng.tau.ac.il}
\thanks{ - Corresponding author. \\This research was supported by the Israel Science Foundation (grant 1503/10)}.
\affiliation{School of Electrical Engineering, Tel Aviv University, Ramat-Aviv, Tel-Aviv 69978  Israel}

\begin{abstract}
Confluent with the single dimension of time, breach of time-reversal symmetry is usually perceived as a one-dimensional concept. In its ultimate realization--the one-way guiding device--it allows optical propagation in one direction, say $+z$, and forbids it in the opposite direction, $-z$. Hence, in studies of time-reversal asymmetry the mapping $t\mapsto -t$ is naturally associated with $z\mapsto -z$. However, strongly non-reciprocal or one-way nano-scale threads can be used \emph{to weave meta-surfaces} thus adding dimensions to this concept. In this new family of surfaces the aforementioned association \emph{cannot be made}. An example of appropriate threads are the planar one-way particle chains based on the two-type rotation principle. The resulting surfaces--the meta-weaves--posses generalized non-reciprocity such as ``sector-way'' propagation, and offer new possibilities for controlling light in thin surfaces. We study several meta-weave designs and their asymmetries in the wave-vector space.

\end{abstract}


\maketitle

\section{Introduction}

Strongly non-reciprocal structures and one-way propagation schemes have attracted considerable attention in the last decade. Numerous different configurations were suggested to create one-way structures, most of them share a common concept. First, one violates Lorentz's reciprocity, either by making the susceptibility $\bb{\chi}$ asymmetric using magnetization or by time modulation of $\bb{\chi}$. Hence propagation in opposing directions possesses different sets of features. Then, another mechanism (e.g.~geometric) is employed to make one set preferable, yielding one-way behavior.
One way total reflection from infinite periodic magneto-optical (MO) layers were demonstrated in \cite{YUFAN_APL}.
In other configurations, one-way behavior exists at the interface between two Photonic Crystals (PhC), or between a PhC and a metal, where at least one of them consists of MO or gyro-magnetic material
\cite{HaldaneRaghu,RaghuHaldane,Soljacic_PRL08,YUFAN,Soljacic_Nature09,FangYuFanPRB_2011,LDS_ChiralHallEdge}. Photonic topological insulators based on edge-states between two bianisotropic metamaterials were suggested in \cite{PTI_2013}. In all these schemes the one-way edge-states are assumed to be completely separated from the surrounding free-space by the semi-infinite \emph{supposedly impenetrable} structures on both sides.
 One-way transmission through a screen-assembly composed of perforated perferct electric conductor (PEC) placed at the interface of a MO media is shown in \cite{SpoofPlasmons}. Likewise, one-way transmission through combined screens of MO material and $\epsilon$-near zero (ENZ) material \cite{OneWay_ENZ}, through combined screens of MO and negative-$\epsilon$ materials \cite{MONegE}, or through a screen of magnetized ENZ in Voigt configuration \cite{MENZ}, were suggested. Non-magnetic one-way behavior was achieved by time-modulation of $\epsilon$ \cite{YuFan_NatPhot,FanLipsonExper_2012}. In all the schemes above, the transverse dimensions of the one-way structure must be of several wavelengths (or several PhC periods) to operate properly.

\begin{figure}[htbp]
        \includegraphics[width=7.1cm]{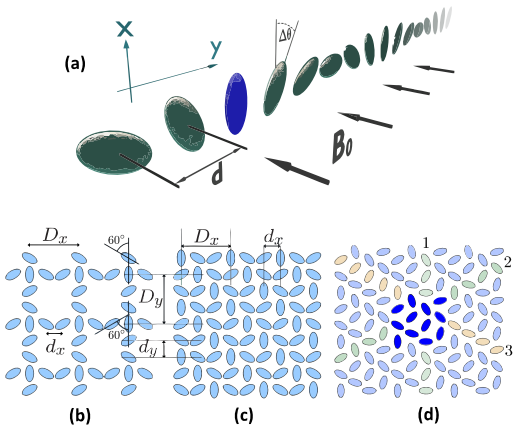}
    \caption{\figstyle{A basic one-way thread that can be used to weave surfaces, and our meta-weaves. (a) A \emph{planar} chain of plasmonic ellipsoids with transverse magnetization and \emph{longitudinal chirality}, supporting one-way guiding. (b)-(d) Some optional weaves, subject to a bias magnetization $\bb{B}_0=\unit{z}B_0$. (b) A ``snug'' rectangular weave of two identical chains with $\Delta\theta=60^\circ$ ($D_x=D_y=3d$). (c) A tight rectangular weave of the same chains. (d) A tight hexagonal weave of any two of the three chains with $\Delta\theta_{1,2,3}=60^\circ, 75^\circ, 15^\circ$. The rectangular period is marked by dark ellipsoids.}}
    \label{fig1}
\end{figure}

One-way guiding structures consisting of a single linear chain of nano-scale plasmonic particles were suggested in \cite{HadadSteinbergPRL,MazorSteinbergPRB,HadadSteinbergNanoAnt,HAMAS}.
These studies include analytical models based on the Discrete Dipole Approximation (DDA), and full-wave simulations with material loss and finite particle size verifying the one-way property for realistic parameters.
The underlying physics is based on the interplay of two types of rotations: geometric and electromagnetic. An examples is shown in Fig.~\ref{fig1}a.
 A chain of plasmonic particles supporting trapped plasmonic modes is exposed to transverse magnetization $\bb{B}_0=\unit{z}B_0$.  $\bb{B}_0$ induces \emph{longitudinal rotation} of the chain modes: the excited dipole in each particle rotates in the $x,y$ plane. Then, a \emph{longitudinal chirality} is introduced by using non-spherical particles that rotate in the $x,y$ plane, with rotation step $\Delta\theta$, as shown in the figure. Two-type rotations coexist in a single plane, and their interplay enhances non-reciprocity and creates one-way guiding \cite{MazorSteinbergPRB}. This structure has several appealing properties. (a) It possesses nano-scale transverse size. (b) Propagation in the ``forbidden" direction decays by two orders of magnitude over distances of $O(\lambda)$. (c) $B_0$ is weaker than other magnetization-based approaches. (d) Since both rotations take place in a single plane that coincides with the chain axis, particle dimension in the $\unit{z}$ direction is unimportant; one may use flat ellipsoidal flakes as particles. Hence, the structure is \emph{flat}, amenable for planar fabrication.

The purpose of the present work is to generalize the one-dimensional concept of broken time-reversal symmetry, by suggesting the \emph{meta-weaves}. Being a nano-scale wide, flat, one-way ``thread,'' the longitudinal-chirality chain in Fig.~\ref{fig1}a may be used to ``weave'' meta-surfaces that \emph{add dimensions} to the \emph{concept} of broken time reversal symmetry. In these structures the natural association of the 1D mapping  $t\mapsto -t$ with the 1D mapping $z\mapsto -z$ \emph{cannot be made}. Rather, the former needs to be associated with a higher dimensional mapping in the plane. To characterize this association and the ensuing optical behavior we define the notion of \emph{sector-way} propagation; a structure is said to be $\phi$ sector-way if, when excited by a point source, it allows propagation only into a cone whose vertex angle is $\phi$. Our meta-weaves possess \emph{sector-way} propagation dynamics and offer new possibilities for controlling the flow of light in a plain.

There are many ways to weave a surface. Some examples are shown in Figs.~\ref{fig1}(b)-(d). The most natural choice is the ``snug'' weave, defined as the case where the weave period (=inter-thread distance) matches the chain period. The ``tight'' weave is defined as the case when the inter-chain distance is the same as the inter-particle distance within the corresponding single chain. The weave periods in the $\unit{x},\unit{y}$ directions $D_x,D_y$ for the snug and tight weaves coincide with the corresponding chains period.

Due to inter-threads coupling, the meta-weave properties may not always be inferred by a mere ``product'' of the single threads. Nevertheless, sector-way propagation is observed even in tight weaves, reflecting the robust nature of the two-type rotation principle and its compatibility for multi-dimensional nonreciprocity.

Our weaves are systematically structured as follows. All particles possess the same shape, differing only by rotation. The $m,n$ \emph{lattice} location $\rv_{m,n}$ and rotation $\theta_{m,n}$, are given by
\begin{eqnarray}
\rv_{m,n} &=& m\bb{a}_1+n\bb{a}_2\label{eq1}\\
\theta_{m,n} &=& m\Delta\theta_1+n\Delta\theta_2.\label{eq2}
\end{eqnarray}
Here $\bb{a}_1,\bb{a}_2$ are the fundamental lattice vectors along which the chains are weaved, and $\Delta\theta_{1,2}$ are the corresponding rotation steps. In tight weaves all lattice points are occupied by a particle, but every non-tight weave possesses empty points. We denote by $\mathbb{P}$ the set of all occupied points.

We use the discrete dipole approximation (DDA) to study our meta-weaves. Under the DDA, a particle response to an exciting local field $\bb{E}^L$ (the field at the particle's location, in the absence of the particle), is described by its dipole moment $\bb{p}=\bb{\alpha}\bb{E}^L$, where $\bb{\alpha}$ is the particle polarizability matrix. It formally holds when the particle size $D_p$ is much smaller than $\lambda$ and when the inter-particle distance $d\gg D_p$. However, studies show excellent agreement with
exact solutions even when $d = 1.5D_p$ \cite{MaierKikAtwater}. Also, full wave simulations with finite particle size and material loss show that the one-way chains dynamics is predicted well by the DDA \cite{MazorSteinbergPRB,HadadSteinbergNanoAnt}. Note that $\bb{B}_0$ affects only the $xy,yx$ entries of $\bb{\alpha}$. Hence the $z$ components of $\bb{\alpha}$ can be ignored, rendering $\bb{\alpha}$ a $2\times 2$ matrix, and $\bb{p}$ a two-elements vector. In our weaves, the $m,n$ particle polarizability $\bb{\alpha}_{m,n}$ is
\begin{equation}
\bb{\alpha}_{m,n} = {\bf T}_{-\theta_{m,n}}\bb{\alpha}{\bf T}_{\theta_{m,n}}\label{eq3}
\end{equation}
 where ${\bf T}_\theta$ is a rotation by $\theta$ operator in the $x,y$ plane, and $\bb{\alpha}$ is the polarizability of a reference ellipsoidal particle. A Drude-model $\bb{\alpha}$ of a magnetized ellipsoid that takes into account the particle's radiation loss is used here (see e.g.~\cite{MazorSteinbergPRB}). We assume lossless material. It has been shown that material loss doesnot change essentially the one-way thread properties if one uses much denser chains, but in this case valid modeling requires full-wave simulations \cite{MazorSteinbergPRB}. For our structures this is beyond currently available computing power. With the definitions above, the surface modes are governed by the difference equation
\begin{equation}\label{eq4}
\bb{p_m}=\bb{\alpha_m}\! \sum_{\bb{\scriptstyle{m'}}\in\mathbb{P}_{\bb{\scriptstyle{m}}}} \!\! {\bf G}(\bb{r_m},\bb{r_{m'}})\, \bb{p_{m'}},\,\, \bb{m}\in\mathbb{P}
\end{equation}
where $\bb{m}$ ($\bb{m}'$) denotes the integers pair $m,n$ ($m',n'$). ${\bf G}(\rv,\rv')$ is the dyadic Green's function \cite{JACKSON}, hence ${\bf G}(\rv,\rv')\bb{p}$ gives the electric field at $\rv$ due to a dipole $\bb{p}$ at $\rv'$. The set $\mathbb{P}_{\bb{\scriptstyle{m}}}$ is the set $\mathbb{P}$ excluding the point $\bb{m}$.

The eigensolutions of \Eq{eq4} constitute the surface modes (i.e. in-plane propagation). It can be reduced to a finite matrix by exploiting the weave periodicity. We find that using the rectangular periodicity is the most convenient approach even for hexagonal weaves, mainly because symmetry-based $k$-space reductions (e.g. irreducible Brillouin zone) cannot be applied due to loss of reciprocity. Furthermore, the rectangular representation is sometimes more efficient even in hexagonal weaves. For example, the parallelogram period of the lattice in Fig.~\ref{fig1}d obtained from the hexagonal lattice vectors consists of 36 particles, while the rectangular periodicity cell consists of only 12 particles. Hence, we define $\mathbb{P}^0$ as the restriction of $\mathbb{P}$ to a reference rectangular period containing the origin. $\mathbb{P}^0$ consists of $M$ particle locations $\rv_1,\ldots\rv_M$, with the corresponding $M$ polarizabilities $\bb{\alpha}_1,\ldots\bb{\alpha}_M$ [\Eq{eq3} consists of at most $M$ different $\bb{\alpha}_{m,n}$'s]. Each point $\bb{m}\in\mathbb{P}$ can be expressed by the three integers $m,\ell_x,\ell_y$ where $m\in\{1,\ldots M\}$ counts the points in $\mathbb{P}^0$, and $\ell_x,\ell_y$ count the unit cells. By periodicity, the dipole response in each particle $\bb{p}_{m,(\ell_x,\ell_y)}$ satisfies
\begin{equation}\label{eq5}
\bb{p}_{m,(\ell_x,\ell_y)}=\bb{p}_me^{i\bb{\beta}\,\cdot\, \bb{D}_{\ell_x,\ell_y}}.
\end{equation}
where $\bb{\beta}=\unit{x}\beta_x+\unit{y}\beta_y$ is the wave-vector, and $\bb{D}_{\ell_x,\ell_y} = \unit{x}\ell_x D_x+\unit{y}\ell_y D_y$.
Using the above in \Eq{eq4}, we obtain a matrix equation with $M\times M$ blocks of $2\times 2$ submatrices, governing the $M$ vectors $\bb{p}_m$,
\begin{equation}\label{eq6}
\left(\bb{\alpha}_m^{-1}-\bb{S}_{m,m}\right)\bb{p}_m -
\sum_{\substack{\scriptscriptstyle{n=1} \\ \scriptscriptstyle{n\neq m} }}^{M}\bb{S}_{m,n}\bb{p}_n=\bb{0}.
\end{equation}
where $\bb{S}_{m,n}$ are the $2\times2$ matrices
\begin{equation}\label{eq7}
\bb{S}_{m,n}=\sum_{\ell_x,\ell_y}\!\!{}^\prime\,
{\bf G}\left(\rv_m,\rv_n+\bb{D}_{\ell_x,\ell_y}\right)e^{i\bb{\beta}\,\cdot\, \bb{D}_{\ell_x,\ell_y}}
\end{equation}
and where the $\ell_x,\ell_y$ summation is over all integers. The prime indicates that the summation excludes the singular self-term arising in $\bb{S}_{m,m}$ when $\ell_x=\ell_y=0$. This summation converges poorly, but it can be accelerated using the Ewald method \cite{EwaldMethod,Ewald2}, modified to account the self term exclusion \cite{Cappolino_PRE_2011}. The dispersion $\omega(\beta_xD_x,\beta_yD_y)$ is obtained numerically by nullifying the corresponding determinant. Since our meta-weaves are \emph{non-Bravais} lattices there are $M$ dispersion surfaces for each of the particle's resonances, that need to be searched.

\begin{figure}[htbp]
    \centering
    \hspace*{-0.2in}
        \includegraphics[width=7cm]{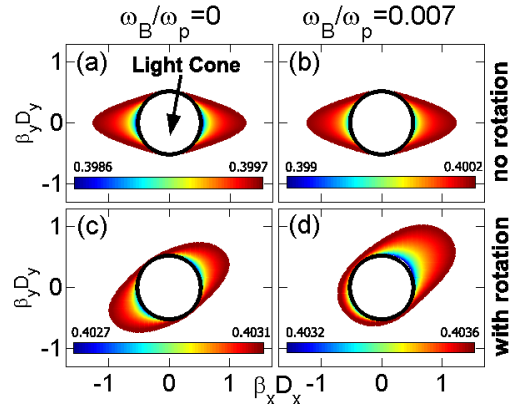}\vspace*{-0.1in}
    \caption{\figstyle{Dispersion surfaces color-coded according to frequency in units of $\omega/\omega_p$. (a) A weave with $\Delta\theta_{1,2}=0$ and $B_0=0$. (b) $\Delta\theta_{1,2}=0$, $B_0\ne 0$. (c) $\Delta\theta_{1,2}=60^\circ$, $B_0=0$. (d) $\Delta\theta_{1,2}=60^\circ$, $B_0\ne 0$.}}
    \label{fig2}
\end{figure}

\begin{figure}[htbp]
    \centering
        \includegraphics[width=7.5cm]{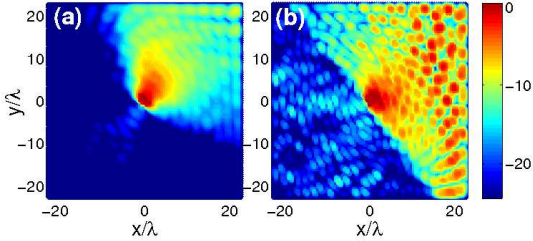}\vspace*{-0.3in}
    \caption{\figstyle{The response $|\bb{p}_{m,n}|$ in dB of the snug weave of Fig.~\ref{fig1}b and Fig.~\ref{fig2}d, to a dipole at the origin. (a) $\sim\pi/2$ sector-way guiding at $\omega=0.40343\omega_p$. Fore Cu parameters, this sector-way is preserved over a bandwidth of more than 100GHz. (b) $\sim\pi$ sector-way at $\omega=0.40355\omega_p$.}}
    \label{fig3}
\end{figure}

We turn now to some examples, starting with the snug rectangular weave of Fig.~\ref{fig1}b. The parameters are $d_x=d_y=\lambda_p/14.5$ ($\lambda_p=2\pi c/\omega_p$). The particle's axis ratios are $a_x:a_y:a_z=1:0.9:0.25$, where $a_x=d_x/4$. With these parameters the DDA is highly accurate. For Cu ($\lambda_p=142$nm) particles diameter is $2a_x\approx 5$nm and $d\approx 10$nm (for larger Cu particles see last example). Also $\Delta\theta=60^\circ$, and $\omega_b=-eB_0/m_e=7\cdot 10^{-3}\omega_p$ denotes magnetization strength (cyclotron frequency). The particle possesses two resonances (associated with $a_x$ and $a_y$), and there are $M=5$ particles in a period. Hence we have $10$ dispersion surfaces, only some of which support sector-way guiding. An example is shown in Fig.~\ref{fig2}.
It is seen that symmetry under the operation $(\beta_xD_x,\beta_yD_y)\mapsto -(\beta_xD_x,\beta_yD_y)$ is broken when rotation and magnetization are simultaneously introduced.
This strong asymmetry leads to ``sector-way'' guiding, explained as follows. There is a well-established theory of these one-way threads \cite{HAMAS}, showing that the excitation magnitude of a thread mode near the light-cone scales as
\begin{equation}\label{eq8}
A=\frac{1-x}{\ln(1-x)},\quad x=\abs{\bb{\beta}}/k_0.
\end{equation}
Hence, modes touching or residing very close to the light cone are practically non-excitable.
Fig.~\ref{fig2}d shows that the dispersion contours within the blue to green range, touch or nearly touch the light-cone in the third quadrant; the normalized distance $1-x\ll 0.05$. Hence \Eq{eq8} predicts a reduction of more than two orders of magnitude in the corresponding modes excitation. Therefor propagation in these directions (given by the dispersion's local gradient) is practically blocked. Figure \ref{fig3} shows the response of this weave to an excitation of a unit dipole at its center. A $\pi/2$ and a $\pi$ sector way propagations are seen. The latter picture exhibits much stronger oscillations than the former. This is because in the $\pi/2$ sector, most of the reflections occurring at the surface edge are in ``forbidden'' directions, hence they decay exponentially as they leak energy to the free space. In the $\pi$ sector-way case some reflections occur at allowed directions and interfere with the modes propagating towards the edge.

\begin{figure}[H]
    \centering
    \hspace*{-0.2in}
        \includegraphics[width=8.5cm]{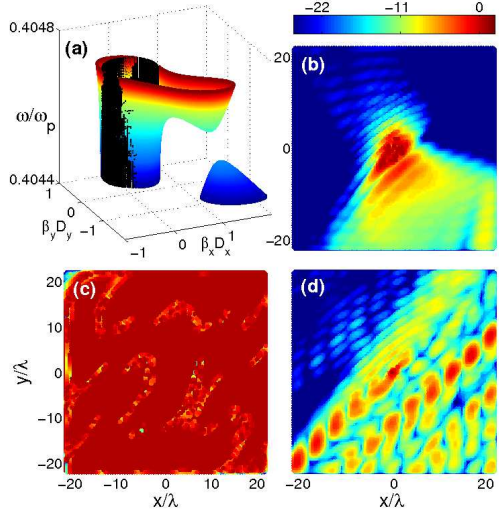}\vspace*{-0.1in}
    \caption{\figstyle{(a) Another dispersion surface of the snug weave of Fig.~\ref{fig1}b. The light cone is shown by the black cylinder in the center. (b)-(d) Responses, (b) at $\omega=0.404585\omega_p$ possessing $\pi/2$ sector-way guiding, (c) at $\omega=0.40463\omega_p$ ``all-way'' guiding, and (d) at $\omega=0.404695\omega_p$ $\pi$-way guiding.}}
    \label{fig4}
\end{figure}

\begin{figure}[htbp]
    \centering
        \includegraphics[width=7.7cm]{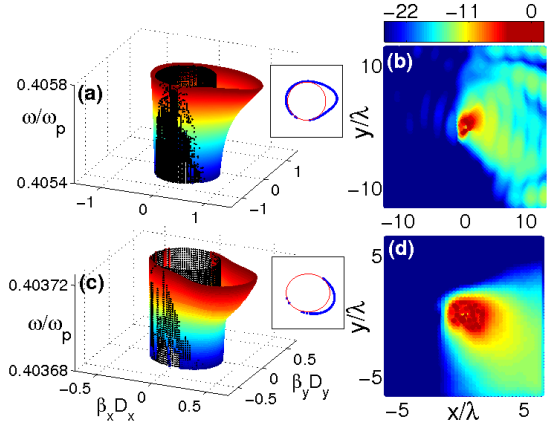}
    \caption{\figstyle{Dispersion and response of the tight rectangular and hexagonal weaves of Fig. \ref{fig1}(c)-(d). (a) Dispersion surface of the tight rectangular weave. (b) Sector-way response at $\omega=0.40561\omega_p$. (c) Dispersion surface of the tight hexagonal weave. (d) Sector-way response at $\omega=0.403685\omega_p$. The insets show the corresponding dispersion contours.}}
    \label{fig5}
\end{figure}

Figure~\ref{fig4} shows another surface of the same weave, and a response to a dipole excitation at three different frequencies. $\pi/2$ sector-way, $\pi$ sector-way, and all-way are observed. The sector-way shown in Fig.~\ref{fig4}b is \emph{not} obtained by an obvious ``cartezian product'' of the individual threads. Such a product would predict a sector that coincides with one of the plane quadrants, while the sector obtained is centered approximately around $-\unit{y}+0.3\unit{x}, \,\, y<0$. The high intensity saturated field in Fg.~\ref{fig4}c is due to the fact that reflections at edges are all into allowed directions, hence they fill the surface and increase the intensity.
 Now to the tight weaves. Figures \ref{fig5}a,b show the dispersion and the response of the tight rectangular weave of Fig.~\ref{fig1}c with the same parameters as above. Figures \ref{fig5}c,d show the dispersion and the response of the tight hexagonal weave of Fig.~\ref{fig1}d with inter-particle distance $d\lambda_p/30$, with $a_x:a_y:a_z=1:0.9:0.25$ where $a_x=d/4$, and with the same magnetization as above. Sector-way propagation is observed in both weaves.

\begin{figure}[htbp]
    \centering
        \includegraphics[width=7.2cm]{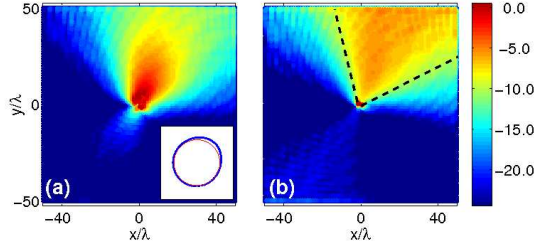}\vspace*{-0.2in}
    \caption{\figstyle{Response of the snug rectangular weave of Fig. \ref{fig1}(b), using larger particles. The inset shows the corresponding dispersion contour. Here $\omega=0.3988\omega_p$ ($\approx 820$THz for Cu). (a) Dipole response $\abs{\bb{p}_{mn}}$. This sector-way is essentially preserved over $\approx 0.8$THz bandwidth. (b) $\sqrt{r}\abs{\bb{p}_{mn}}$, i.e.~the same response, but without the geometrical spreading effect. Along the dashed lines the intensity is reduced by $e^{-2}$.}}
    \label{fig6}
\end{figure}

As a last example, Fig.~\ref{fig6} shows the response of the snug rectangular weave of Fig.~\ref{fig1}(b), using larger Cu particles with $2a_x=16$nm and the same axis ratios as before. Here $d=24$nm. Sector way is observed. Finally, to get a better feeling of the nature of the meta-weave trapped modes, Fig.~\ref{fig6}(b) shows the same solution but multiplied by $r^{1/2}$ where $r$ is the distance from the source at the center. This clears out a $r^{-1/2}$ decay due to 2D geometrical spreading. It is seen that now there is no decay at all along the sector central line. The same behavior applies to all the previous examples.

To conclude, a new family of meta-surfaces, the meta-weaves, were suggested and studied. These meta-weaves are made of strongly non reciprocal or one-way threads based on the two-type rotation principle. It has been shown that they suggest a systematic generalization of the one-dimensional concept of broken time-reversal symmetry, and its extension to higher dimensions. The result is a surface that exhibit sector-way guiding features that may offer new ways to control the flow of light in thin surfaces.

\end{document}